# Cyclic plasticity and fatigue damage of CrMnFeCoNi high entropy alloy fabricated by laser powder-bed fusion

Minsoo Jin[1], Bogdan Dovgyy[1], Ehsan Hosseini[2], Alessandro Piglione[1], Paul. A. Hooper[3], Stuart. R. Holdsworth[2,4], Minh-Son Pham[1]

1. Department of Materials, Imperial College London, Exhibition Road, South Kensington, London SW7 2AZ, UK
2. EMPA, Swiss Federal Laboratories for Materials Science and Technology, Überlandstrasse 129, 8600 Dübendorf, Switzerland
3. Department of Mechanical Engineering, Imperial College London, Exhibition Road, South Kensington, London SW7 2AZ, UK
4. Inspire Centre for Mechanical Integrity, c/o EMPA, 8600 Dübendorf, Switzerland

## Abstract

The CrMnFeCoNi high-entropy alloy is highly printable and holds great potential for structural applications. However, no significant discussions on cyclic plasticity and fatigue damage in previous studies. This study provides significant insights into the link between print processes, solidification microstructure, cyclic plasticity and fatigue damage evolution in the alloy fabricated by laser powder bed fusion. Thermodynamics-based predictions (validated by scanning transmission electron microscopy (STEM) energy dispersive X-ray spectroscopy (EDX)) showed that Cr, Co and Fe partition to the core of the solidification cells, whilst Mn and Ni to the cell boundaries in all considered print parameters. Both dislocation slip and deformation twinning were found to be responsible for plastic deformation under monotonic loading. However, the former was found to be the single dominant mechanism for cyclic plasticity. The surface finish helped to substantially delay the crack initiation and cause lack-of-fusion porosity to be the main source of crack initiation. Most significantly, the scan strategies significantly affect grain arrangements and grain dimensions, leading to noticeable effects on fatigue crack propagation; in particular, the highest resistance crack propagation was seen in the meander scan strategy with 0° rotation thanks to the most columnar grains and the smallest spacing of grain boundaries along the crack propagation path.

## 1. Introduction

Laser Powder-Bed Fusion (LPBF) is an additive manufacturing (AM) method that involves layer-by-layer fabrication using a computer-controlled power source to consolidate a material into a desired shape. The interest in AM alloys is rising rapidly, since AM methods enable the fabrication of components with complex and optimised geometries, reducing fabrication time and costs. Successful use of any alloys in structural application requires understanding of long-term mechanical performance of the alloys. However, studies on the long-term performance of AM alloys are still limited.

High-Entropy Alloys (HEA) are multicomponent alloys that consist of 5 or more principal elements in (near) equiatomic concentration, first introduced by Cantor *et al.* [1] and Yeh *et al.* [2] in 2004. The term 'HEA' was introduced to signify the high configurational entropy of mixing, which is significantly larger compared to that of conventional alloys [2–4]. Although it is a matter of debate, the high mixing entropy of the HEAs is believed to stabilise solid-solution phases with a higher configurational entropy (typically body-centred cubic, face-centred cubic or hexagonal closed packing system), and hinder the formation of intermetallic compounds with a lower configurational entropy [2,3]. Among many types of HEAs, the equiatomic CrMnFeCoNi alloy is one of the most studied ones, as it exhibits a single face-centred cubic (FCC) solid-solution phase (as first reported by Cantor *et al.* [1]). Although the presence of Cr-rich precipitates has been reported in this HEA [5–7], the alloy remains of great interest due to its remarkable mechanical behaviour under monotonic loading over a wide range of temperatures, holding great potential for structural applications in cryogenic conditions [8–11].

The CrMnFeCoNi HEA is reported to be highly printable via LPBF [12,13]. Such printability is partly due to its narrow freezing range (60 °C), which lowers the alloy's susceptibility to solidification/liquation cracking [14]. Considering the alloy's great potential for structural applications, AM would enable a wider realisation of such potential. In addition to many studies on HEAs manufactured by casting [6–10,15], there is an increasing number of studies on as-printed microstructures and the associated mechanical properties of the CrMnFeCoNi HEA fabricated by AM [12,13,16]. However, the majority of such studies only focus on the alloy's monotonic tensile properties, with no significant understanding of cyclic plasticity and the associated fatigue behaviour. Thus, in this study, process defects and solidification microstructure of the CrMnFeCoNi HEA fabricated by LPBF are examined. Fatigue tests were carried out to study cyclic plasticity and establish links between deformation behaviour, porosity and solidification microstructures. In particular, this study reveals a considerable effect of the print strategy on the fatigue crack propagation of the alloy.

## 2. Experimental Procedures

*2.1 Materials and Powder-bed selective laser fusion*

The CrMnFeCoNi samples were manufactured via LPBF using a Renishaw AM250 printer equipped with a modulated wave fibre laser on a 316L stainless steel substrate in argon atmosphere. Pre-alloyed CrMnFeCoNi powder was provided by H.C. Starck Surface Technology & Ceramic Powders GmbH. The samples were printed using a laser power of 200W, a hatch spacing of 85 μm, an exposure time of 80 μs, a point distance of 60 μm, a layer height of 50 μm and a spot size of 65 μm. These printing parameters were selected following the optimized print parameters for the 316L stainless steel (which is of the FeCrNi alloy system as the HEA) and those used for CrMnFeCoNi samples in previous studies [12,17].

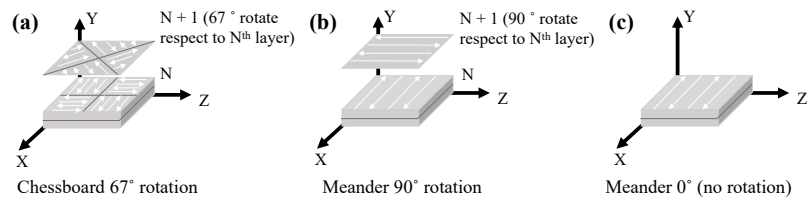

Figure 1. Schematics of the print strategies used in this study: (a) chessboard scanning pattern - each layer is subdivided into 4x4mm islands (within which the laser follows a meander pattern, which is rotated by 90° with respect to the neighbouring island) with 67° rotation between layers. (b) meander scanning pattern with 90° rotation between layers. (c) meander scanning pattern with no rotation between layers.

Cylindrical dog-bone fatigue samples with a gauge length of 18 mm and a diameter of 4 mm were manufactured with the build direction parallel to their longitudinal axis. Three distinct scan strategies were employed to study their effect on the alloy's fatigue performance, Figure 1. Furthermore, some builds were manufactured with the same scan strategy (i.e. chessboard pattern) but different surface conditions (as-printed surface or machined surface) to investigate the effect of surface finish and sub-surface porosity. The relative densities of the dog-bone samples were calculated using Archimedes' Principle.

*2.2 Mechanical testing and damage evolution analysis*

The fatigue tests were performed at 22 °C under strain control at ±0.5 % strain amplitude (R=-1) with strain rate of $10^{-2}$ s$^{-1}$. Fatigue life was defined at a number of cycles when the stress response dropped by 20 % from a plateau region. The yield stress in cyclic loading was evaluated at an offset strain of 0.01%. The damage evolution in the fatigue specimens was investigated by analysing the difference between apparent elastic moduli in tensile-going loading ($E_t$) and in compression-going loading ($E_c$). In general, fatigue cracks, if existed, should remain open in the beginning of compression-going loading from the tensile peak stress due to the tensile stress state, but remain closed at the beginning of the tensile-going loading due to the compressive stress state. Therefore, if there are sufficiently long cracks or pores in the specimen, $E_t$ should be higher than $E_c$, as shown in Figure 2. The difference between the apparent elastic moduli in tension and compression ($\Delta E_{t-c}$) becomes larger as cracks propagate; in this study, such difference is plotted against the number of cycles and fitted with a function of the form $10^{(aN+b)}$ to quantify the crack propagation behaviour (similar to the method used in [18]). The coefficient *a* indicates the rate of change of $\Delta E_{t-c}$: the faster the cracks advance in each cycle, the larger the *a* coefficient. The value of *b* indicates the initiation of fatigue cracks: the higher the *b* coefficient (closer to 0), the earlier the initiation. It is shown later in this study that fatigue crack initiation is associated with porosity in the alloy. Therefore, the *b* coefficient also reflects the porosity condition of as-built samples. In addition to the fatigue tests, monotonic tensile tests were performed using an Instron 300DX at 22 °C with a strain rate of $10^{-3}$ s$^{-1}$ on samples that were printed horizontally using a chessboard pattern with a 67° rotation between layers.

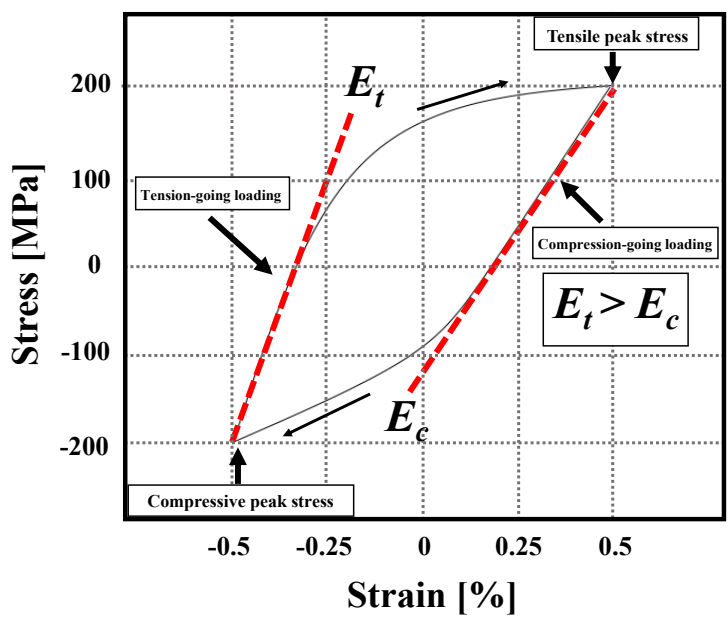

Figure 2. Cyclic loading hysteresis loop with the elastic modulus of the tension-going loading ($E_t$) from the compressive peak stress and that of the compression-going loading ($E_c$) from the tensile peak stress. The apparent elastic moduli difference between $E_t$ and $E_c$ are due to crack opening and closing during cyclic loading.

*2.3 Microstructure characterisation*

Fracture surfaces were observed with a Zeiss™ Auriga scanning electron microscope (SEM). Samples for electron backscattered diffraction (EBSD) were prepared by using 1200, 2400, and 4000 grit SiC papers followed by polishing in an OP-S solution with 0.4 μm colloidal silica particles. Polished samples were further electrochemically etched in 10 % oxalic acid solution for 30 s at 10 V. EBSD analyses were carried out using a Zeiss™ Sigma 300 SEM equipped with a Field Emission Gun (FEG) electron source. EBSD analyses were performed using a step size of 0.1 μm per pixel; grain boundaries were defined when a misorientation larger than 5° was detected. The MTEX toolbox, together with the dislocation density estimation function by Pantleon [19], was used to analyse geometrically necessary dislocation (GND) density evolution [20]. The step size of EBSD data used for the GND density estimation was 0.3 μm per pixel. Specimens for Transmission Electron Microscopy (TEM) analysis were prepared either by electrochemical polishing using a twin-jet Struers Tenupol and an electrolyte consisting of 10% perchloric acid in a methanol at 20 V and -20 °C, or by focused ion beam (FIB) lift-outs using a Helios NanoLab 600. Bright Field (BF) TEM and scanning TEM (STEM) were performed using a JEOL 2100F field emission TEM operating at 200 kV. An Oxford Instruments energy dispersive X-Ray spectroscopy (EDX) system with INCA software was used for chemical analysis. Theoretical solidification behaviour was analysed using calculated phase diagram methods (CALPHAD) using ThermoCalc 2019a with a step of 1 K and the TCHEA2 thermodynamics database. The Scheil solidification module was used to approximate the rapid cooling of LPBF.

## 3. Results

*3.1 Consolidation*

Table 1 shows that the HEA had an excellent consolidation: samples fabricated by the chessboard 67° scan and the meander 90° scan with the as-printed surface condition had a relative density of 99.6 %, while a sample fabricated by the meander 0° scan with the as-printed surface had a slightly lower relative density (99.2 %). The relative density of the machined chessboard 67° scan sample was higher than that of the chessboard 67° sample with the as-printed surface, indicating that machining removed the as-built porosity in sub-surface regions.

Table 1. Relative densities of the samples fabricated by different print strategies, measured using Archimedes' Principle.

| Print strategies | Relative density of sample (%) |
| --- | --- |
| Machined Chessboard 67° | 99.8 |
| As-printed Chessboard 67° | 99.6 |
| As-printed Meander 90 ° | 99.6 |
| As-printed Meander 0 ° | 99.2 |

*3.2 As-printed microstructure*

EBSD maps coloured according to the inverse pole figure (IPF) along the build direction (BD) and the grain size distributions corresponding to the considered print strategies (Figure 3) showed that the meander 0° scan had the smallest grain size. The mean and median of grain size distribution for the meander 0° sample were 14.5 µm and 6.8 µm, respectively. The grain size distribution for the chessboard 67° and the meander 90° samples showed high similarity; the mean and the median were respectively 27.4 µm and 17.1 µm for the chessboard 67° sample and 26.8 µm and 17.4 µm for the meander 90° sample. The aspect ratio of grains (Figure 4) was calculated as the ratio of their longest dimension to their shortest dimension, as detected in the EBSD maps (Figure 3 a, c and e), to reflect their columnarity. The cumulative distribution of the grains aspect ratio for as-printed samples (Figure 4) showed that the aspect ratios of grains associated with the meander 0° and the meander 90° scans were similar and higher than that of the chessboard 67° scan, indicating that grains induced by the two former scans were more columnar than those of the latter one.

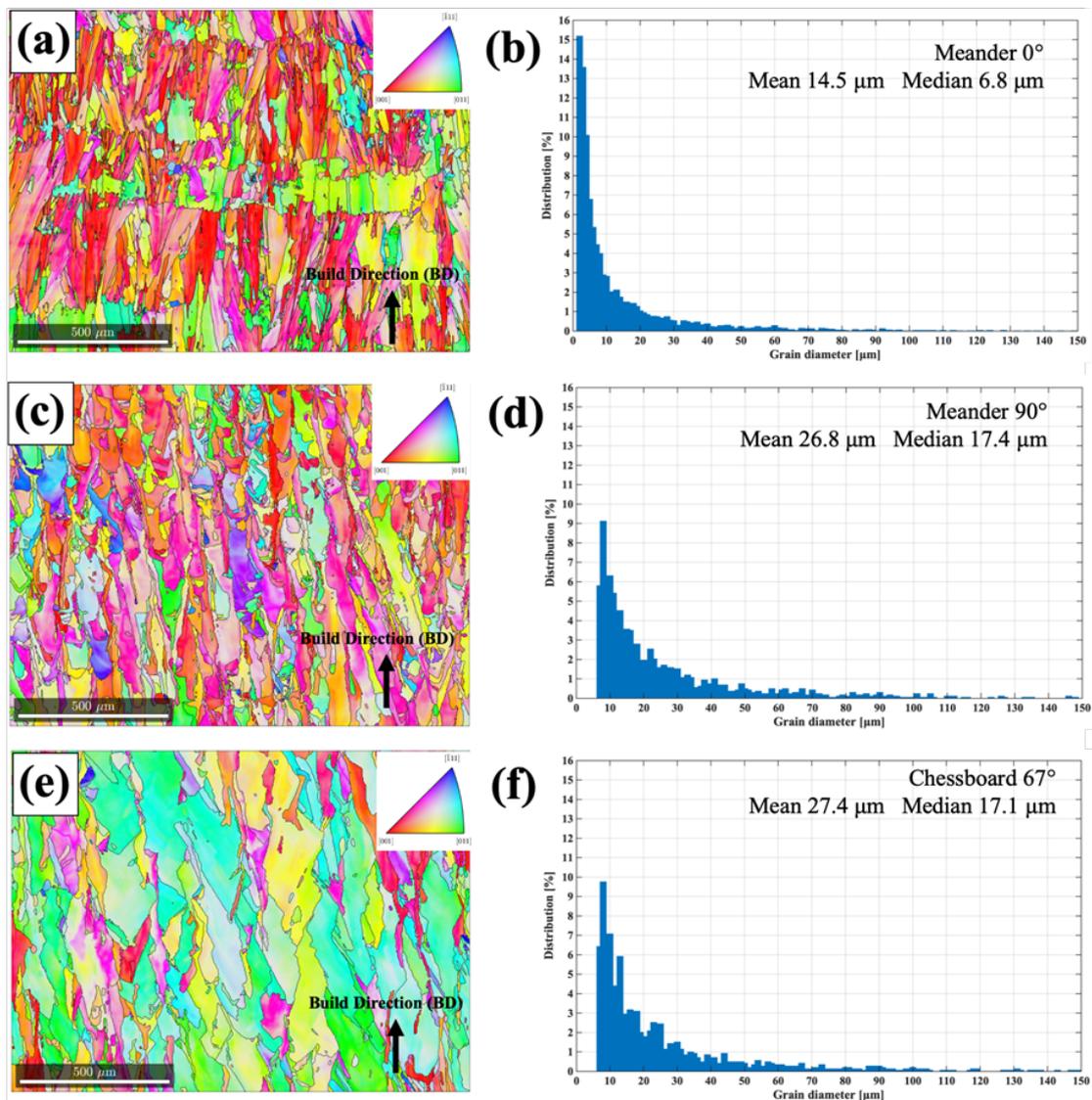

Figure 3. EBSD IPF-BD maps and grain size distributions of (a), (b) the meander 0° scan (no rotation), (c), (d) the meander 90° scan and (e), (f) the chessboard 67° scan.

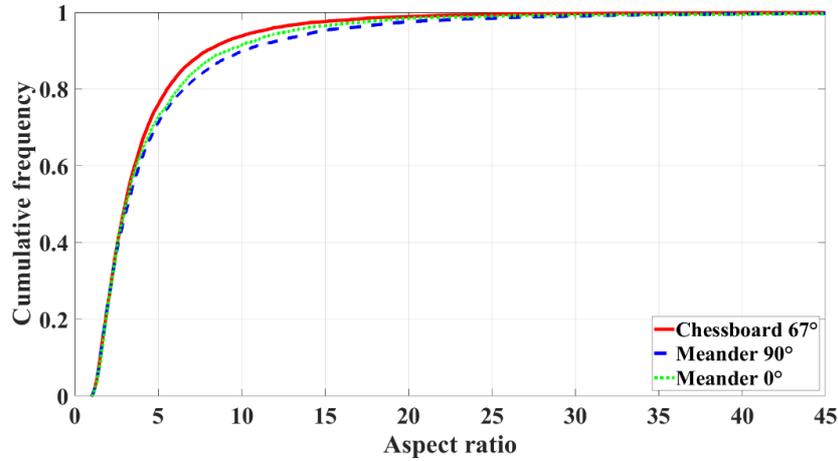

Figure 4. Cumulative distribution of grains aspect ratio for the as-printed samples fabricated by the different print strategies.

All samples exhibited fine cellular microstructures, regardless of print strategies. It is believed that high cooling rates (~$10^6$ K/s) of LPBF [21] are responsible for the formation of fine cells and/or dendrites in LPBF alloys [22–25]. The appearance of cell boundaries, as revealed by electrochemical etching (Figure 5a), indicates the presence of elemental segregation towards cell boundaries. Figure 5b shows a simulation of solidification behaviour of the CrMnFeCoNi HEA, obtained using the Scheil-Gulliver calculation. Liquidus and solidus temperatures of this alloy were identified to be about 1600 K and 1430 K, respectively. During solidification, the root (core) of cells forms first (at the liquidus temperature), and then grows radially towards cell boundaries. Therefore, in Figure 5b the mass fraction of solid 0.0 represents the core of cells, whilst 1.0 represents the cell boundaries. CALPHAD simulations showed that Ni and Mn segregate to cell boundaries, while Cr, Co and Fe segregate to the cell core. This segregation behaviour is different to that observed by Cantor *et al.* [1]. In their work, Cr and Mn segregated to the inter-dendritic regions, while Ni, Fe and Co segregated to the dendrite core [1]. To validate the CALPHAD prediction, a STEM-EDX analysis was carried out. The STEM dark field (DF) image in Figure 5c revealed that Cr, Co and Fe segregated to the cell core, whereas Mn (and Ni, although to a lesser extent) segregated to the cell boundaries, confirming the CALPHAD prediction, and in an good agreement with previous studies on the same alloy done byLaurent-Brocq *et al.* [26] and Haase *et al.* [27]. STEM-EDX analysis also detected the presence of secondary particles (with a size of 50~100 nm); these were found to be manganese oxides, as they contained high Mn and O concentrations (Figure 2c), in good agreement with other studies [7,16,27].

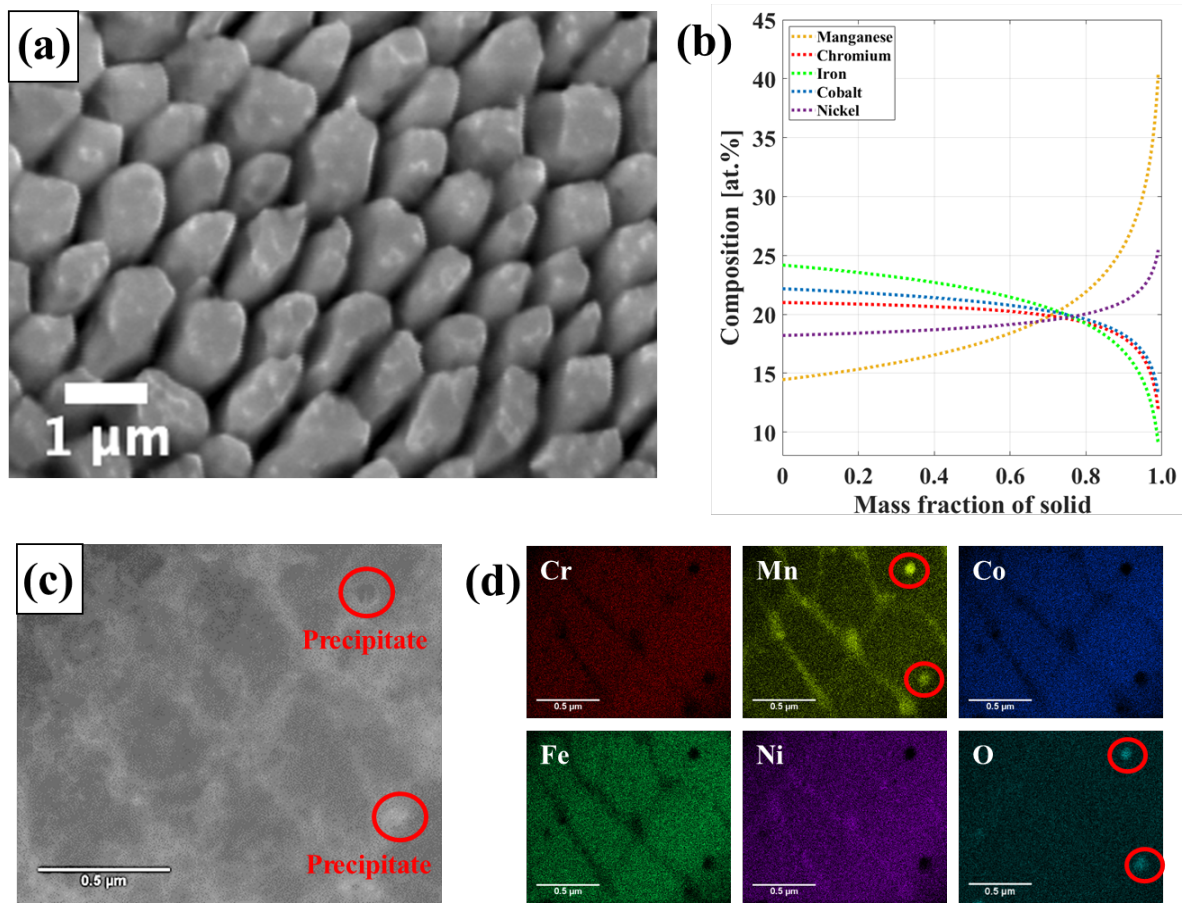

Figure 5. (a) SEM image of the as-printed cellular microstructure in the CrMnFeCoNi HEA after electro-chemical etching, (b) Scheil solidification simulation results of the high entropy alloy performed within Thermo-Calc, (c) STEM-DF image of a region containing several solidification cells, (d) STEM-EDX elemental maps of the region shown in (c) – the locations of the manganese oxides are highlighted.

*3.3 Microstructural evolution*

This section focuses on the evolution of solidification microstructure generated by the same chessboard scan after either monotonic or cyclic loading (Figure 6). TEM examinations showed that dense dislocation tangles were present at solidification cell boundaries in the as-printed condition (Figure 6a). In addition, high resolution EBSD mapping of this microstructure condition (Figure 6b) revealed cell-shaped colour contrasts with a size ranging from 0.5 to 1 μm, which is similar to the size of the cell cross-section, suggesting that such cell-shaped contrast likely resulted from crystallographic misorientation between cells. Such misorientation was measured to be up to 2°. TEM images after cyclic deformation (Figure 6c) show that the cellular microstructure was preserved throughout the cyclic loading. However, the dislocation density within cell boundaries appears higher after cyclic loading. Deformation twinning was not detected in TEM investigation on the fatigued sample; however, it was detected in EBSD maps on the same sample, conducted on larger areas. In particular, Figure 6d shows that {111} deformation twinning ($\sum 3$) was active in the grains that had their [103] orientation parallel to the loading direction. However, deformation twinning was only observed to a limited extent and scattered in regions within 70 μm from fracture surface, explaining why the TEM studies alone did not detect the presence of twinning. In contrast, deformation twinning (∑3) was much more dominant after monotonic loading in <01$\bar{1}$> oriented grains with respect to the loading direction (Figure 6e and f).

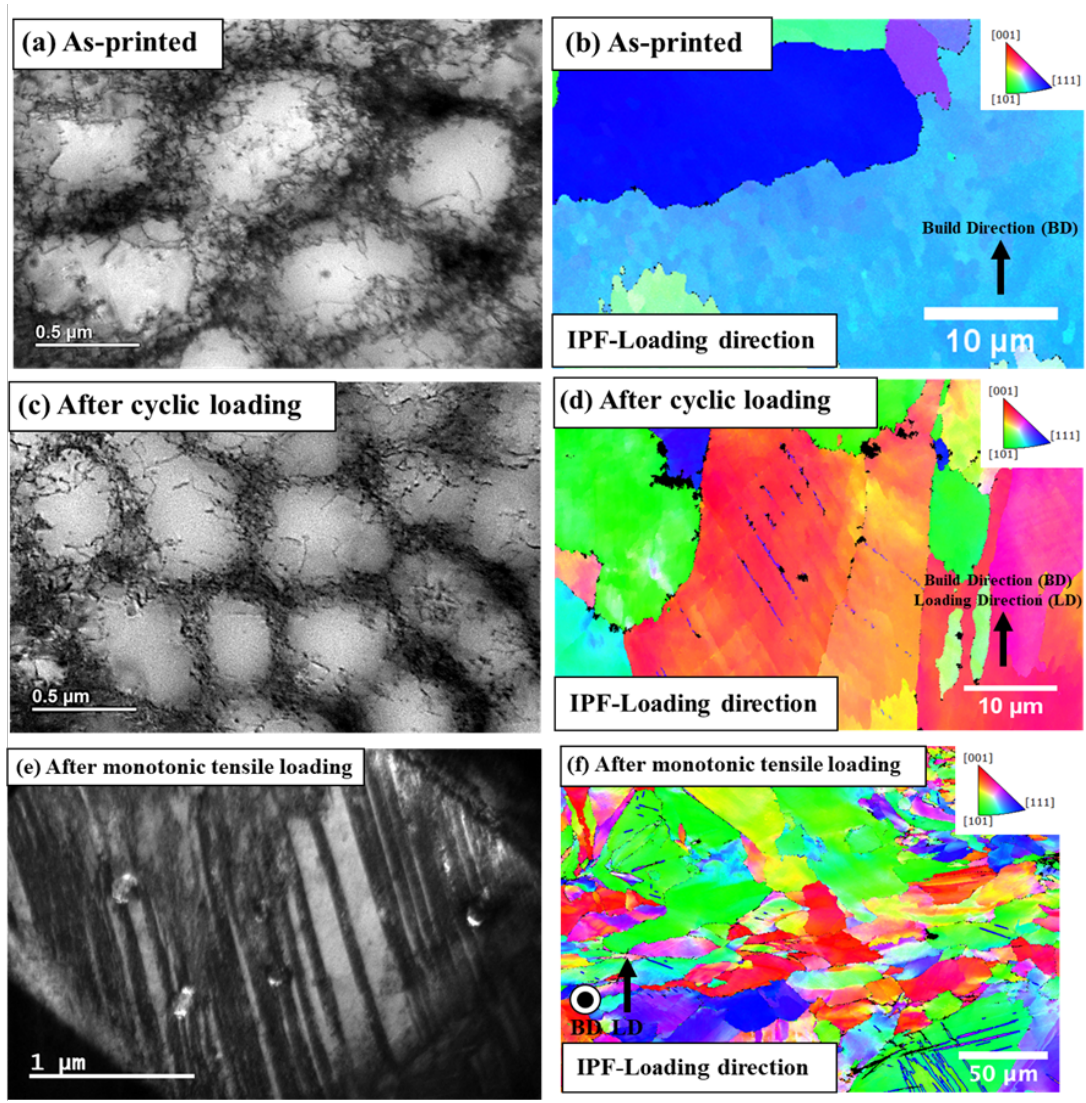

Figure 6. Microstructures of the CrMnFeCoNi HEA fabricated by the chessboard scanning pattern with 67° rotation (a) TEM-BF image of the as-printed condition and (b) EBSD IPF-Build Direction of the as-printed CrMnFeCoNi, where the sample was sectioned parallel to the build direction, (c) TEM-BF image of the CrMnFeCoNi after cyclic loading and (d) EBSD IPF-Build Direction of cyclically deformed CrMnFeCoNi, where the sample was sectioned parallel to the build direction. (e) and (f) TEM-BF image and EBSD IPF-Build Direction of the CrMnFeCoNi after monotonic loading; this TEM foil was lifted from a necked region.

Geometrically necessary dislocation (GND) density estimations based on EBSD mapping (Figure 7 a and b) show that the GND density increased after cyclic loading. However, the increase in GND density was mainly confined to the region within 70 μm from the fracture surface. It was found that the average GND density in this region was 50% higher compared to that outside of this region. On the other hand, the microstructure was heavily deformed after monotonic loading; indeed, Figure 7c shows that the increase in GND density after monotonic loading was much more significant compared to that after cyclic loading (Figure 7b).

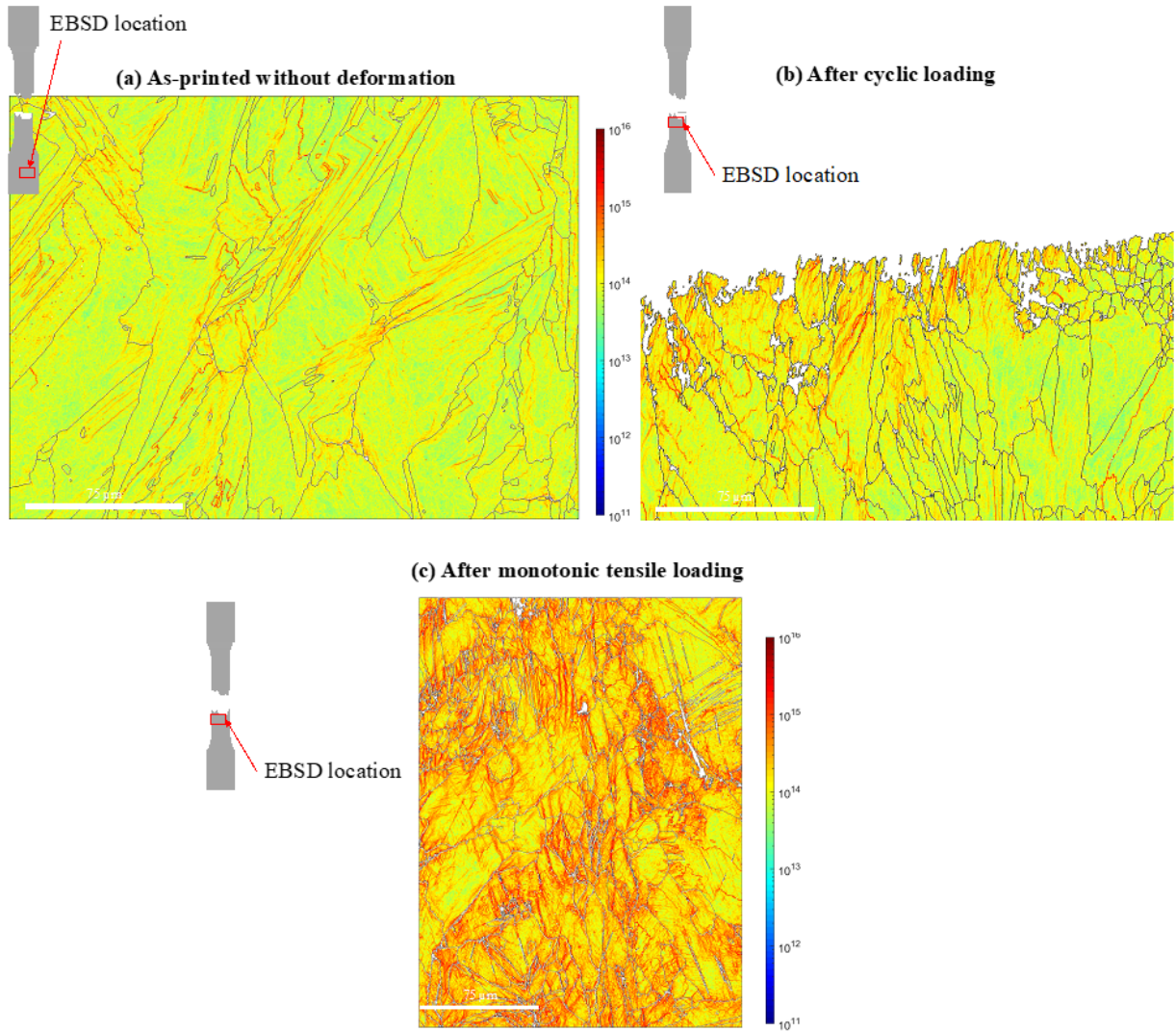

Figure 7. Estimated density of geometrically necessary dislocations of the CrMnFeCoNi HEA fabricated by the chessboard scanning pattern with 67° rotation (a) at thread regions of the sample without any deformation, (b) near the fracture surface after cyclic loading, and (c) near the necking region after monotonic tensile loading. Note: rectangles in the insets highlight locations of the EBSD mapping.

*3.4 Plastic deformation in monotonic and cyclic loading*

The HEA fabricated by LPBF exhibited notable strength and ductility. For example, Figure 8 is stress-strain curve obtained from a monotonic tensile test of a sample horizontally fabricated by the chessboard 67° scan. The yield stress, ultimate tensile strength, and elongation at failure were $520 \pm 10$ MPa, $770 \pm$ MPa and 25%, respectively. Table 2 includes mechanical properties and average grain sizes both from this study and from studies in the literature on the same alloy, fabricated by either vacuum induction melting (VIM) [15] or vacuum arc melting (VAR) [11]. Despite the larger average grain size, the yield stress of the CrMnFeCoNi manufactured by LPBF ($520 \pm 10$ MPa) was superior to both that of the alloy fabricated by VIM ($270 \pm 10$ MPa) and by VAR ($360 \pm 10$ MPa). On the other hand, the elongation at failure decreased slightly by 5-10 % compared to the alloy made using the other processes.

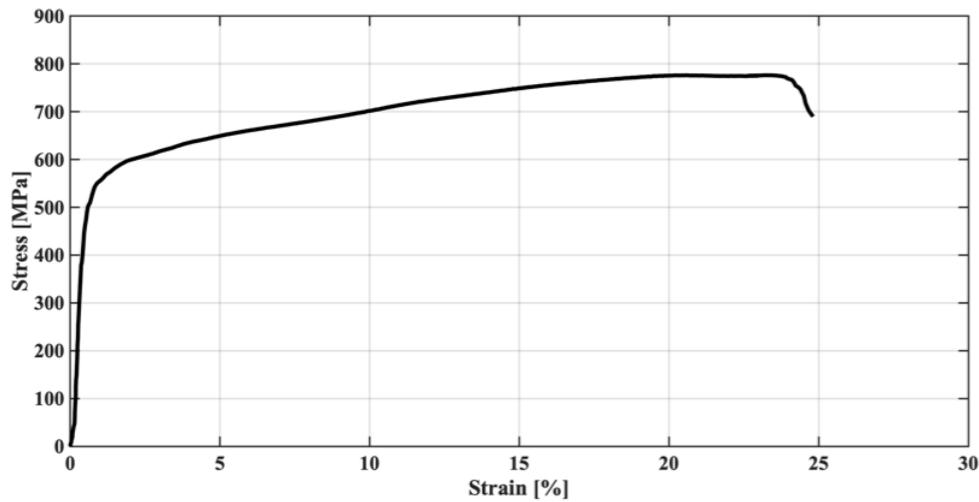

Figure 8. Uniaxial monotonic stress-strain curve of the HEA fabricated by LPBF with the chessboard 67° scan strategy with machined surface condition.

Table 2. Mechanical properties and average grain size (measured from the EBSD maps of the specimen used for monotonic loading) of the CrMnFeCoNi high-entropy alloy at room temperature (293-296 K) manufactured by vacuum induction melting (VIM), vacuum arc melting (VAR) and LPBF.

| Manufacture method | Yield stress (MPa) | Ultimate tensile stress, UTS (MPa) | Elongation at failure (%) | Average grain size (μm) |
|---|---|---|---|---|
| **Laser Powder-Bed Fusion (LPBF) [This Study]** | 520 ± 10 MPa | 770 ± 10 MPa | 25 | 30.0 |
| Laser Powder-Bed Fusion (LPBF) [13] | 520 ± 10 MPa | 600 ± 10 MPa | 28 ± 5 | Not Specified |
| Vacuum Induction Melting (VIM) [9] | 270 ± 10 MPa | 600 ± 40 MPa | 30 ± 5 | 17.0 |
| Vacuum Arc Melting (VAR) [11] | 360 ± 10 MPa | 660 ± 10 MPa | 40 ± 5 | 4.4 |

Figure 9 shows the maximum tensile stress responses of the HEA fabricated by the considered scan strategies during cyclic loading. All samples exhibited a short cyclic hardening behaviour during the first 5 cycles, followed by cyclic softening. The print strategies appear to have some influence on the cyclic plastic deformation, but a negligible effect on the fatigue life of the samples. In contrast, the removal of the sub-surface porosity resulted in a significant improvement in the fatigue performance of a machined sample fabricated with the chessboard 67° scan, as shown by an increase in maximum stress response and by a longer fatigue life.

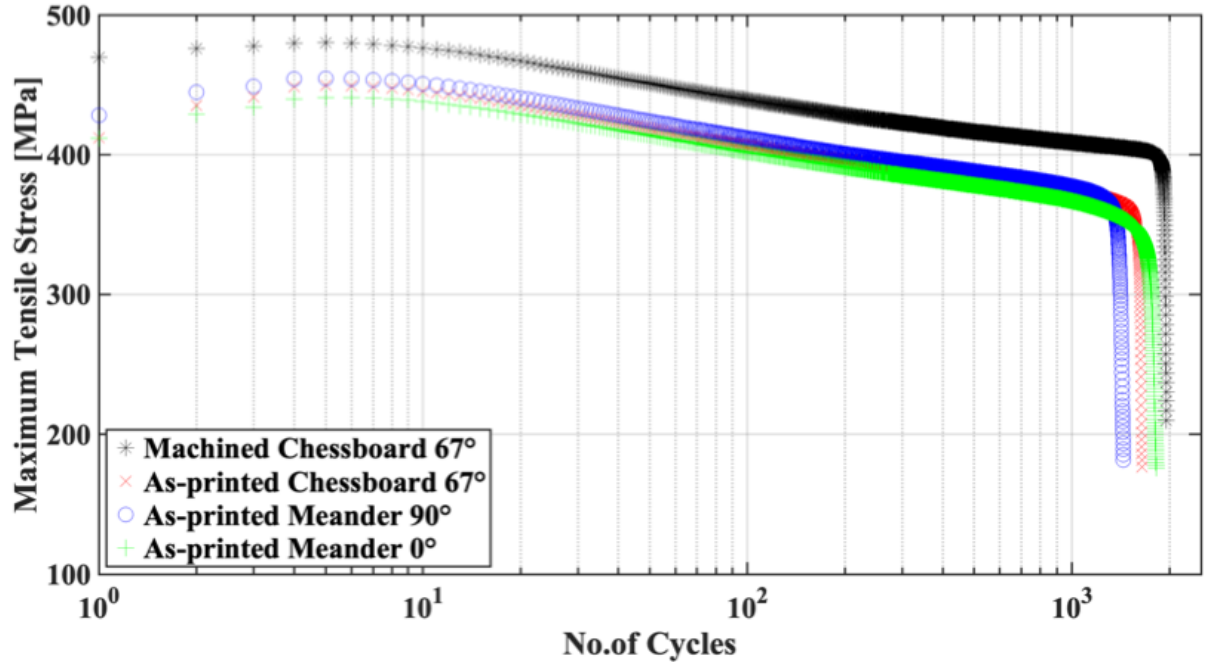

Figure 9. Cyclic plastic deformation response of CrMnFeCoNi HEA samples fabricated using different printing strategies and with different surface conditions.

Figure 10 shows the evolution of apparent yield stresses during cyclic loading of both as-printed and machined samples fabricated by the chessboard 67° scan. During cyclic hardening (i.e. the first 5 cycles) the yield stresses slightly increased, then decreased gradually until failure. The compression-going yield stresses of both samples showed a similar trend during cyclic loading. The difference between the tension-going and compression-going yield stresses became larger (reaching >20 MPa) during the last 50 cycles for both surface conditions. Machining the sample surface resulted in a higher apparent yield stress (by 45 ± 5 MPa on average) compared to that of the as-printed sample, Figure 10a. The different scan strategies used, however, did not have a significant effect on the yield stress evolution during cyclic loading (Figure 10b).

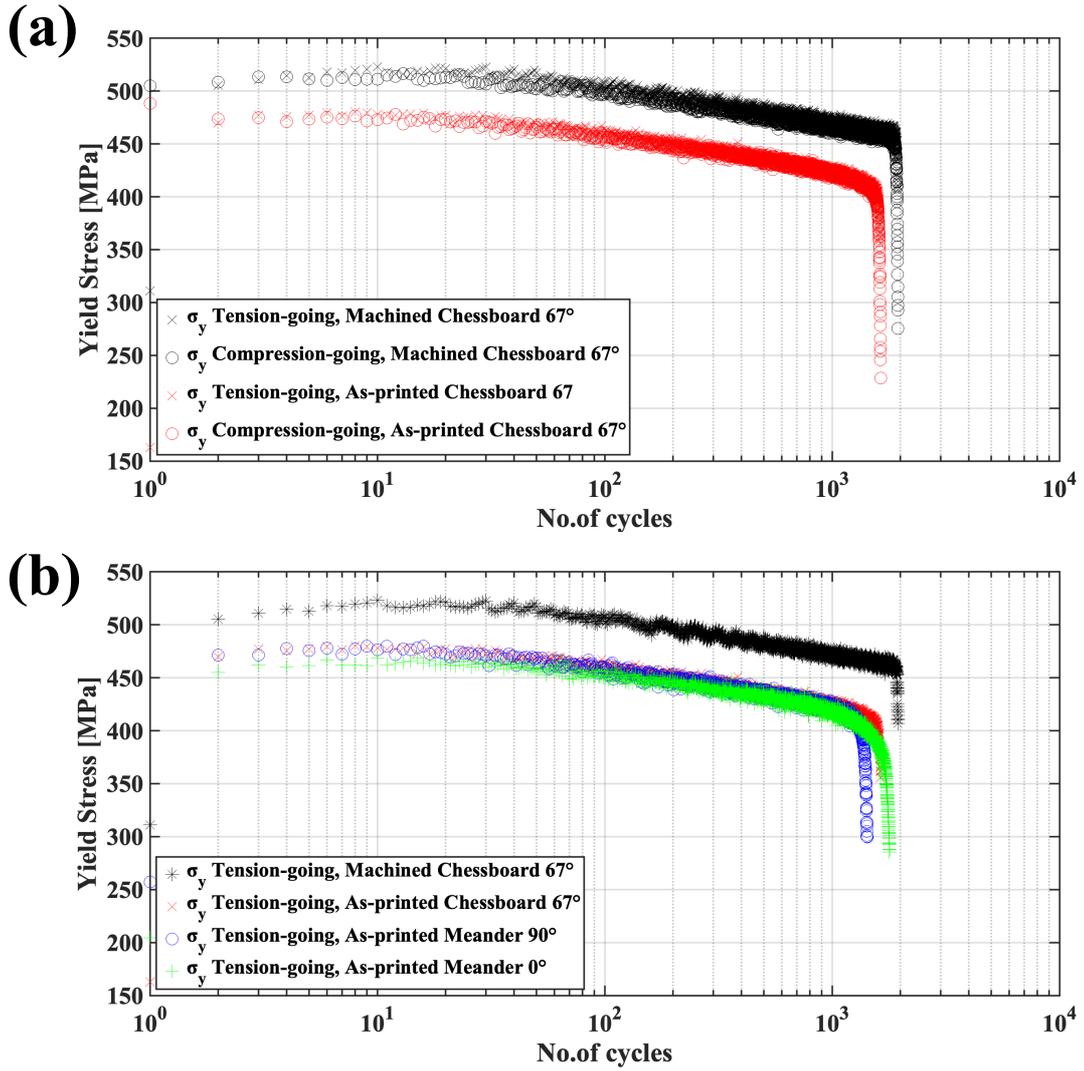

Figure 10. (a) Evolution of the yield stresses of the as-printed and machined chessboard 67° samples during cyclic loading, (b) evolution of the yield stresses of the all samples during tension-going loading.

*3.5 Damage evolution during cyclic loading*

Figure 11a shows the damage evolution of samples fabricated using the same scan strategy (chessboard 67°) but tested with different surface conditions (as-printed versus machined surface). The apparent elastic modulus during the tension-going loading ($E_t$) was higher than during the compression-going loading ($E_c$) by ~ 6 GPa consistently throughout the fatigue life; however, the apparent elastic modulus difference ($\Delta E_{t-c}$) rapidly increased in the last 200 cycles for both surface conditions. In addition, Figure 11a shows that both $E_t$ and $E_c$ of the machined sample were higher than those of the as-printed sample ~15 GPa throughout the entire test. Figure 11b shows the evolution of $\Delta E_{t-c}$ in the last 200 cycles for the two samples, together with the fitting function described in section 2.2. The coefficient *a* for the as-printed and machined conditions were 0.028 and 0.035, respectively. The significantly lower *b* coefficient (-67.5) for the machined sample suggests that fatigue cracks initiated much later compared to the as-printed sample, which had a higher *b* coeffficient of -44.1.

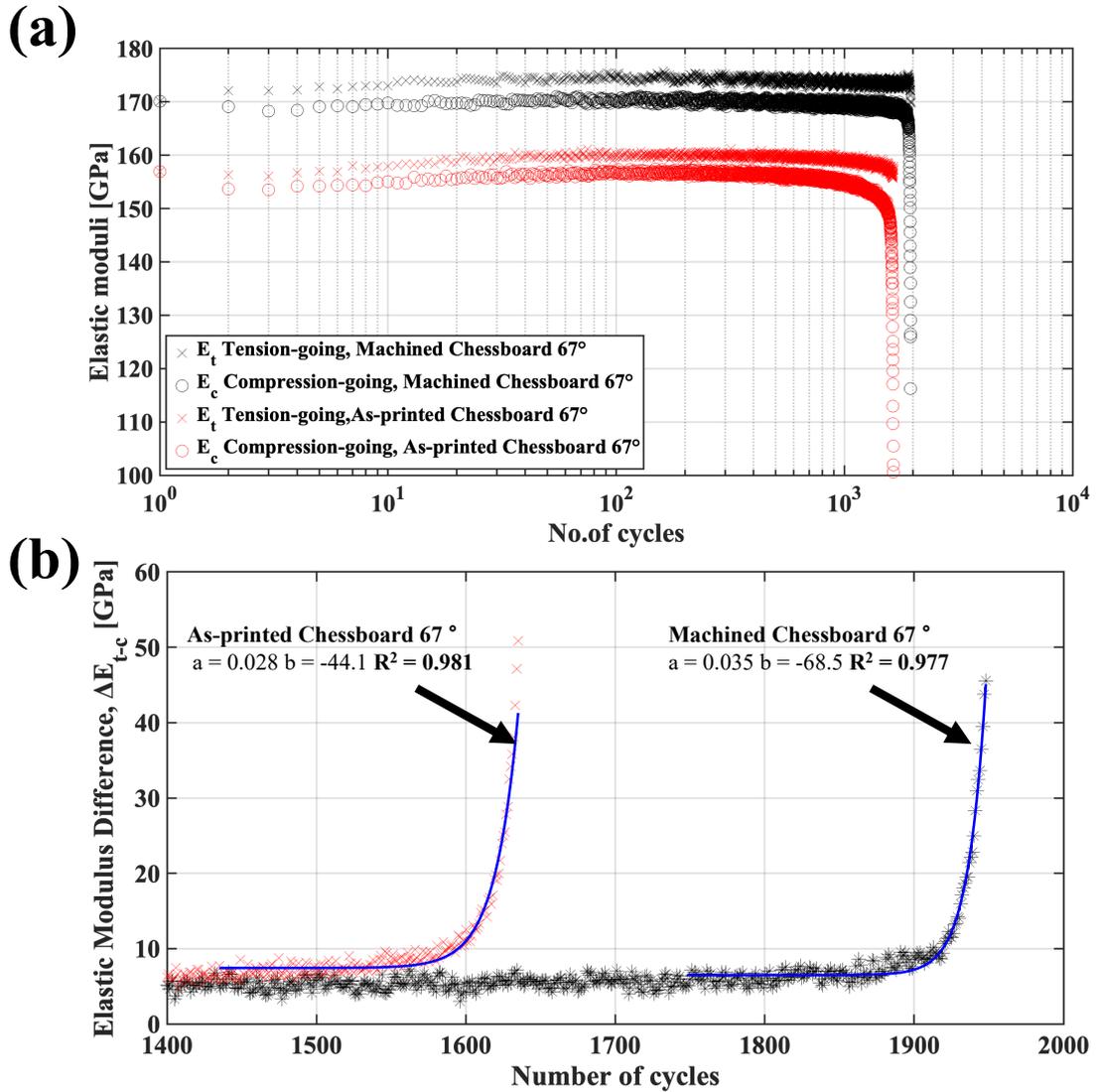

Figure 11. (a) Evolution of apparent elastic moduli during the tensile and the compressive cycles of the as-printed and the machined samples manufactured by the chessboard print strategy with 67° rotation. (b) The difference between the tensile and the compressive elastic moduli for the two samples in (a) with fitted curves

       Figure 12 a and b and Figure 13 show the damage evolution in fatigue specimens fabricated with different print strategies, but with the same as-printed surface condition. Similar to Figure 11a, $E_t$ is consistently larger than $E_c$ regardless of the print strategies (Figure 12). The average $\Delta E_{t-c}$ of all print strategies was about 6 to 10 GPa until the last 200 cycles (Figure 13). The value of the *a* coefficient for the chessboard 67° scan and the meander 90° scan was 0.028. The *a* coefficient for the meander 0° scan was 0.016; this significantly lower value indicates that fatigue cracks advanced slower in this case compared to the other print strategies. The difference in *b* coefficient of the three print strategies was noticeable. The *b* coefficient of the meander 0° scan was -26.6 and was the highest amongst the three scans (-38.7 for the meander 90° scan and -44.1 for the chessboard 67° scan), indicating that crack initiation in the meander 0° scan occurred earlier compared to the other two scans.

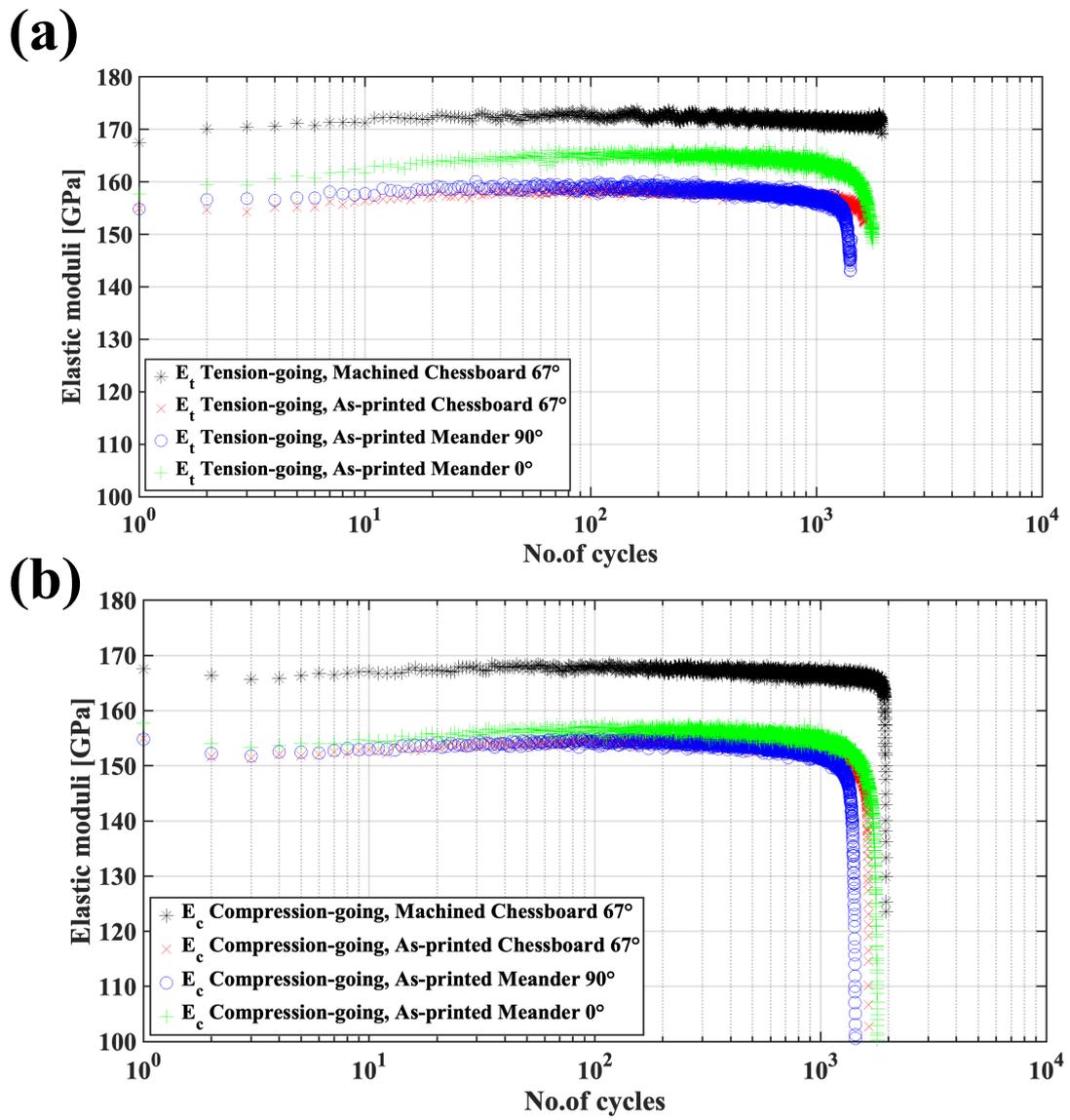

Figure 12. Evolution of the apparent elastic modulus for the samples fabricated with different print strategies (a) during the reverse tensile loading from the compressive peak and (b) from the compressive loading tensile peak stress.

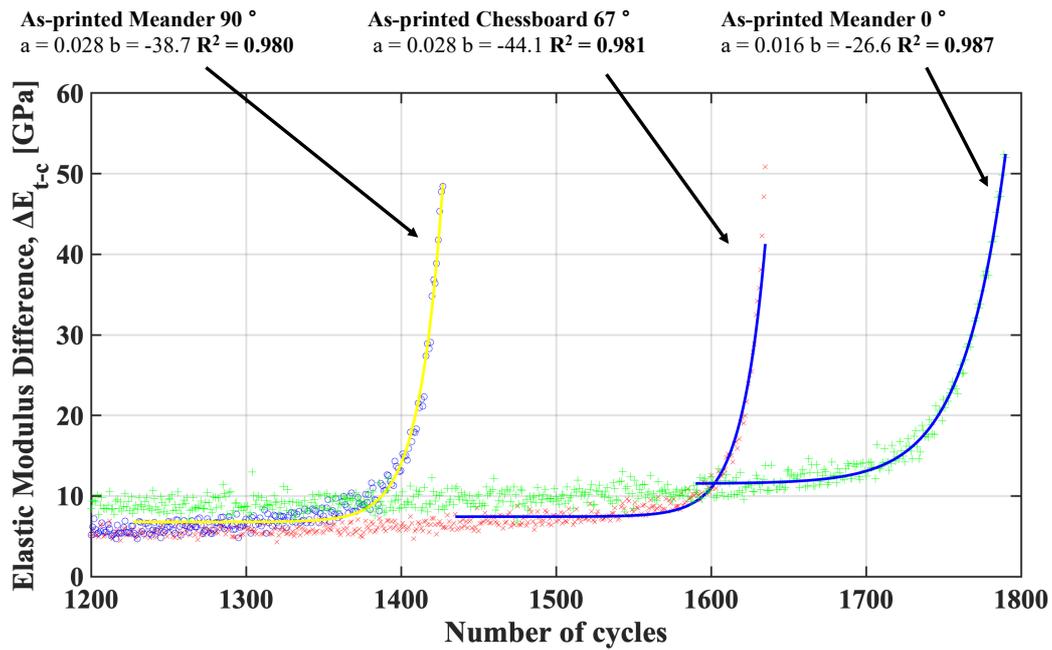

Figure 13. Evolution of the apparent elastic modulus difference for the samples fabricated by the different print strategies in the as-printed condition.

*3.6 Fatigue lives*

Fatigue lives of the HEA samples were calculated as the number of cycles at the 20% drop in the maximum stress response and were summarised in
Table **3**. As previously mentioned, the print strategies appear to have little influence on fatigue lives. The surface condition, on the other hand, appeared to play a more significant role on the fatigue performance; indeed, the fatigue life of the machined sample was improved by about 20% with respect to that of the as-printed surface condition.

Table 3. Fatigue lives of the HEA samples manufactured with different print strategies and surface conditions.

| Surface condition | Print Strategies | Number of cycles at 20% stress drop |
|---|---|---|
| Machined | Chessboard 67° | 1940 |
| As-printed | Chessboard 67° | 1620 |
| As-printed | Meander 90 ° | 1410 |
| As-printed | Meander 0 ° | 1770 |

*3.7 Fracture surface examination*

Dimples were seen on fracture surfaces in regions of overload, corresponding to the end of fatigue life. The presence of dimples indicates that ductile fracture occurred in the LPBF HEA, e.g. Figure 14 a and b (region A of Figure 14a), which is consistent with the good ductility observed during monotonic loading (Section 3.4). Figure 14b is a higher magnification image of ductile dimples, showing small spherical particles embedded within the dimples (Figure 14b), in good agreement with Qiu *et al.*'s observations [16]. These particles are likely to be the Mn oxides observed in the as-printed condition by STEM-EDX mapping (Figure 5b).

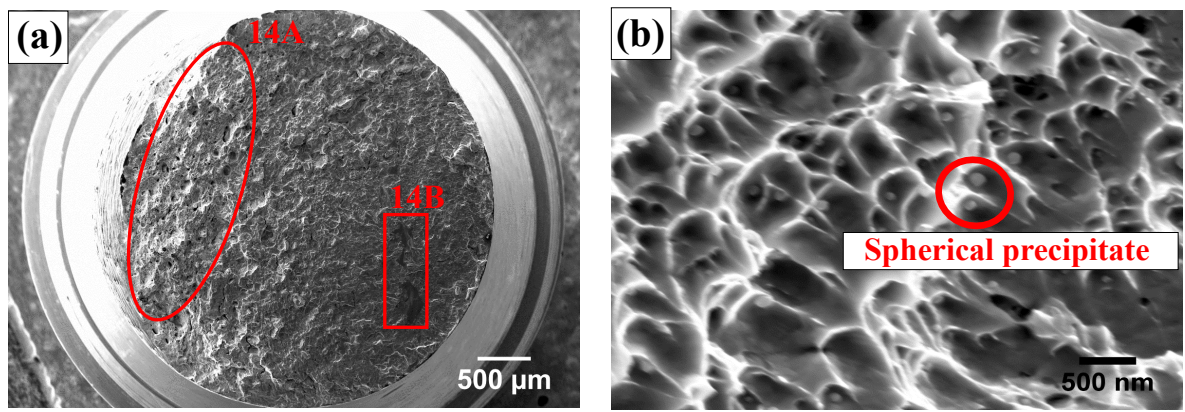

Figure 14. (a) Fracture surface of the sample fabricated by the chessboard 67° scan after cyclic loading; (b) ductile dimples and embedded spherical particles in regions of overload, highlighted as 14A in (a).

Figure 15a shows two large lack-of-fusion (LOF) pores (in the region 14B of Figure 14a), denoted as 15A and 15B; these were observed in close proximity to one another on the fracture surface of the sample with the machined surface. Each of the pre-existing pores had a length of ~400 µm. Their proximity to each other suggests that they possibly merged during cyclic loading to create one large crack with a size of ~800 µm. River-like patterns and fatigue striations radiate from these pores, indicating that they were responsible for the initiation of the dominant crack, which propagated to failure (Figure 15b and c). No evidence of porosity in the sub-surface regions is seen in the machined sample, while all of the as-printed samples had high levels of porosity in the sub-surface regions (Figure 16); no large pores were observed in the bulk.

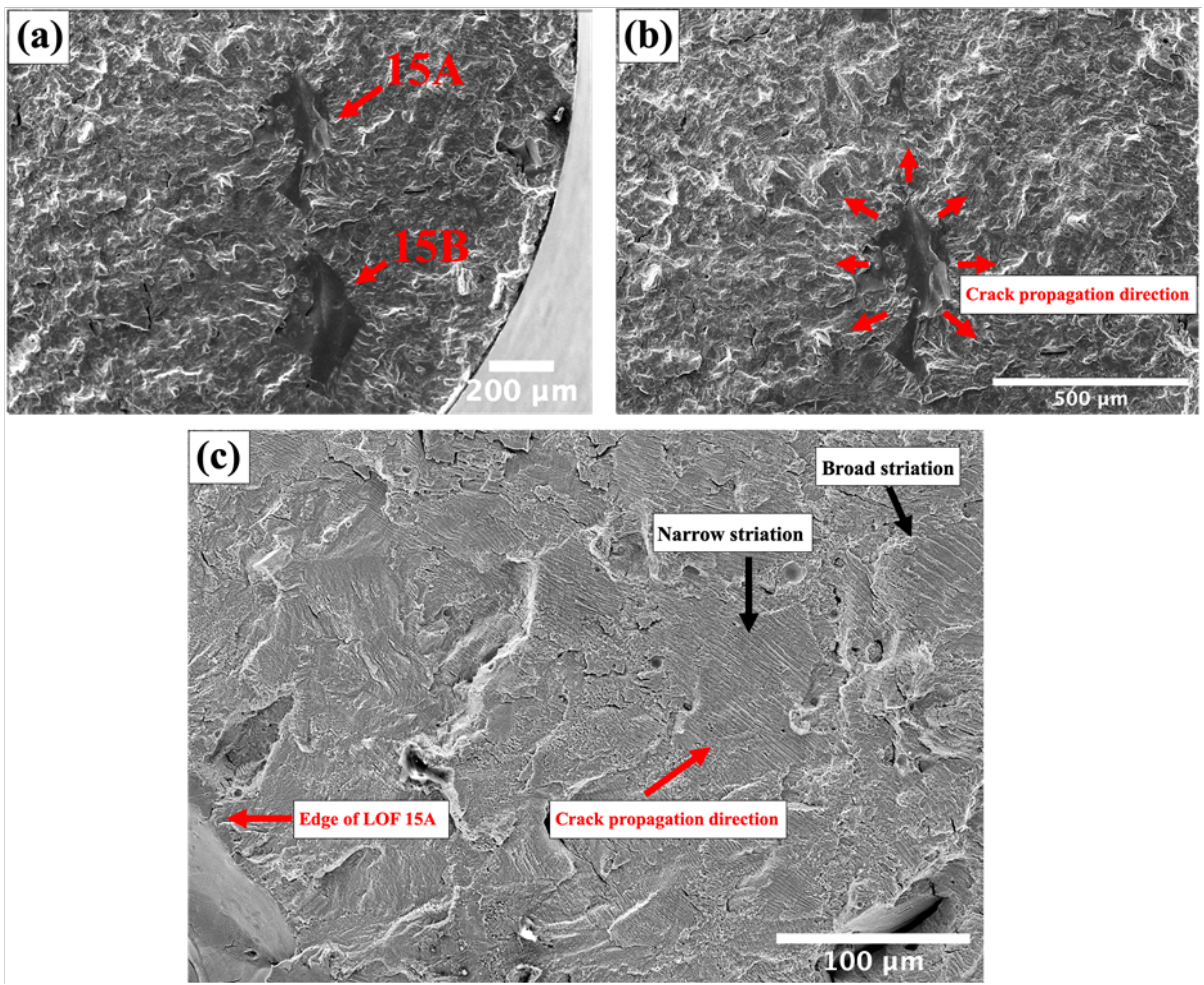

Figure 15. SEM fractographies of the machined sample using the chessboard print strategy with 67° rotation showing (a) the two lack of fusion (LOF) pores denoted as 15A and 15B, (b) river like patterns and propagation directions of the fatigue cracks initiated from the LOF pore 15A, (c) fatigue striations generated by the crack propagation near the LOF pore 15A.

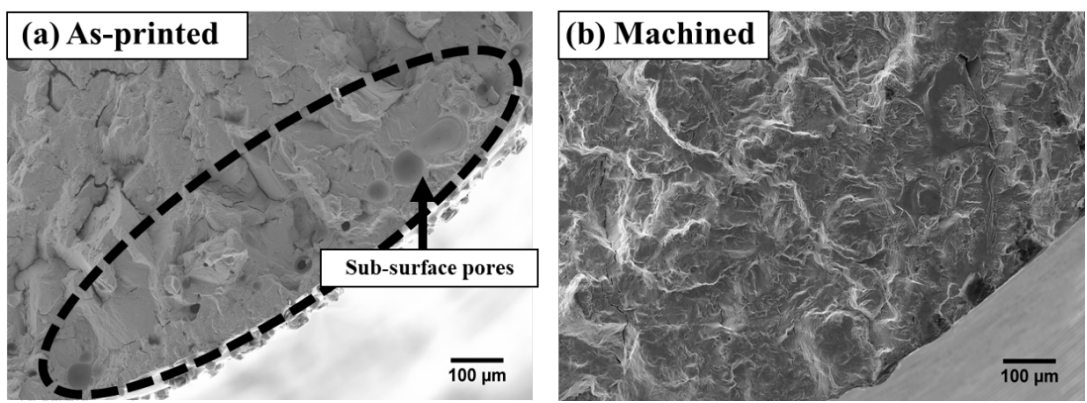

Figure 16. SEM fractographies of the CrMnFeCoNi HEA after the cyclic loading: (a) as-printed chessboard 67° sample showing a line of key-hole pores in the sub-surface region and (b) no evidence of porosity in the sub-surface regions of the machined chessboard 67° sample.

# 4. Discussion

*4.1 Microstructure, tensile strength and cyclic plasticity*

Results in section 3.2 show that the monotonic yield stress of the HEA manufactured via LPBF is superior to that of the same alloy manufactured by VIM or VAR [9,11], despite the larger average grain size. The superior strength is attributed to fine solidification cells (whose boundaries consist of dense dislocations generated by high thermal gradients during cooling) present inside the grains (Figure 6 a and c). Immobile dislocations within cell boundaries can restrict slip of mobile dislocations during deformation, leading to the higher strength of the LPBF HEA. The significant role of solidification cells on the yield strength of AM alloys is also reported for the 316L stainless steel fabricated via LPBF [28].

In FCC metals, deformation twinning is dominant when the stacking fault energy (SFE) is at 15-50 mJ/m$^2$ [29]. The SFE of this HEA is estimated to be 18-27 mJ/m$^2$ at room temperature due to the high Mn concentration (20 at.%) [30,31]. This relatively low SFE promotes deformation twinning during plastic deformation. Otto *et al.* [11] and Laplanche *et al.* [15] also reported deformation twinning in the HEA at room temperature (293 K). However, deformation twinning in the HEA made by VIM and casting is seen only at large strains (usually above 20%), as a minimum critical stress is required for twinning (about 720 MPa) [15]. In contrast, the high initial yield stress of the LPBF CrMnFeCoNi alloy enables it to reach the critical stress (720 MPa) earlier, at a strain of about 11% (Figure 8); this promotes earlier deformation twinning during monotonic loading, as reported in Figure 6f. The occurrence of deformation twinning can reduce the mean free path of dislocations, hence contributing to strain hardening [17,28]. Together with a significant increase in dislocation densities during monotonic loading (Figure 7c), deformation twinning is responsible for the superior ultimate tensile strength of the LPBF HEA as compared to VIM or VAR.

In contrast with substantial changes in microstructure during monotonic loading, there was no significant difference in the alloy's microstructure before and after cyclic loading (Figure 6 a and c), apart from regions near the fracture surface (Figure 7). This was due to the small imposed plastic strain per cycle. *In-situ* TEM investigations of the CrMnFeCoNi HEA [32] showed that plasticity in front of crack tip promotes deformation twinning, which can deflect the propagating crack [32]. In this study, the EBSD analysis (Figure 6d) also showed that {111} deformation twinning ($\sum 3$) was active during cyclic loading in grains with the <103> orientation parallel to the loading direction. However, deformation twinning was only found occasionally and confined to regions within ~70 μm from the fracture surface (Figure 6 d and e). During cracking opening, the stress concentration induced by the crack tip can cause substantial plastic deformation near the fracture surface. Such a higher degree of plastic deformation results in higher dislocation densities (Figure 7) and promoting more twinning near the fracture surface. However, the contribution from deformation twinning to the cyclic deformation behaviour of the alloy would not be significant, as twinning activities were quite limited and very confined.

During cyclic loading, dislocations move to and fro, resulting in more organised dislocation structures in cell boundaries to reduce the stored strain energy [33,34]. In general, the cyclic hardening of annealed alloys is associated with a substantial increase in the dislocation density at the beginning of cyclic loading. At the end of the cyclic hardening, dislocations rearrange themselves into dislocation cell and channel substructures to reduce the stored dislocation strain energy [18,35]. However, in the LPBF CrMnFeCoNi HEA, high thermal gradients and rapid cooling during solidification generated numerous fine cells, whose boundaries contained dense dislocation tangles. This explains a short cyclic hardening (confined to the first 5 cycles) observed in this study. Upon further cyclic loading, mobile dislocations move to and fro, promoting dislocation annihilation within cell interiors while increasing the dislocation density at cell boundaries, resulting in more organised cells, as seen in Figure 6a and c, similar to the evolution of dislocation structures in well annealed conditions [18,36].

*4.2 Influence of processing defects on mechanical properties and fatigue damage*

In fully dense materials with a good surface finish, fatigue crack initiation is often associated with persistent slip bands (PSBs) as a consequence of the to and fro movement of dislocations during cyclic loading [37]. As the cyclic loading continues, plastic strain accumulates and localises at PSBs. Localisation of plastic strain results in extrusions (and intrusions) on the free surface, leading to surface cracks initiating along the interfaces between PSBs and the matrix [37,38]. Therefore, a number of cycles is required to initiate the cracks. Hence, the tension-going elastic modulus ($E_t$) should be similar to the compression-going elastic modulus ($E_c$) in the early stages of cyclic loading. However, the $E_t$ of the LPBF HEA was higher than the $E_c$ from the first cycle for all the samples in this study, regardless of the print strategy used and of the surface condition. This is due to the porosity generated during the fabrication process, and in particular to the lack of fusion pores and key-hole pores shown in section 3.5. Pores affect the stiffness of the tested specimens, and hence the apparent elastic modulus. In the tension-going loading from the compressive peak, the stress state remains compressive during most of the elastic deformation, making existing pores more closed. In contrast, the pores are more open during the elastic deformation of the compression-going loading from the tensile peak, since the stress state is still tensile. Hence, the compressive stress state in the tension-going loading alleviates the negative effect of porosity on stiffness by closing the pores, leading to a higher $E_t$ compared to $E_c$, as shown in Figure 12. A higher porosity density should lead to a larger difference between the two moduli, i.e. to a larger $\Delta E_{t-c}$. The consolidation of the meander 0° sample was the lowest amongst the specimens in this study (Table 1), explaining why its $\Delta E_{t-c}$ was the largest (Figure 13). The identified values of the *b* coefficient indicate that fatigue crack initiation in the meander 0° build occurred much earlier compared to the other strategies; indeed, Figure 13 shows that the $\Delta E_{t-c}$ of this scan strategy deviated slightly very early (consistent with the lowest relative density reported in Table 1).

Figure 15 shows that process pores can act as preferential sites for fatigue crack initiation. In particular, pores with irregular shapes are likely to induce severe stress concentrations, and hence earlier crack initiation. Complex melt-pool dynamics, in particular when the laser beam turns by 180° near the free surface of a build, can promote the formation of key-hole pores in the sub-surface regions [39], explaining the frequent observation of key-hole pores in the sub-surface regions of all the as-printed HEA samples of this study (Figure 16a). Surface machining removed the sub-surface porosity; therefore, no sub-surface pores were observed in the machined specimen (Figure 16b). The removal of sub-surface porosity delayed fatigue crack initiation (Figure 11b), leading to the longer fatigue life of the machined specimen compared to that of the as-printed sample (

Table **3**). The negative impact of the sub-surface key-hole porosity was also evident on the apparent elastic moduli (Figure 11). Such porosity induces a reduction in stiffness, hence the apparent moduli of the as-printed samples were lower than that of the machined sample (Figure 11). In turn, the lower stiffness explains why the measured yield stresses of the as-printed samples were lower compared to those of the machined sample (Figure 10).

*4.3 Influence of print strategies on fatigue damage evolution*

Figure 9 shows that all three considered print strategies had little influence on the fatigue lives of the HEA. However, the fatigue crack propagation appears to be dependent on the scan strategy adopted (Figure 12 and Figure 13). The gradient of the damage evolution reflecting the rate of crack propagation is represented by the value of the *a* coefficient. Despite being the most porous sample, the meander 0° specimen with the as-printed surface was found to have the lowest value of the *a* coefficient, and hence the slowest crack propagation. Figure 3 and Figure 4 show that the meander 0° scan resulted in the most columnar microstructure, with the columnar grains being elongated along the build direction and the loading direction. Therefore, the meander 0° had the highest number of grain boundaries that were perpendicular to the propagation of main cracks. In addition, the rotation between layers in the other scan strategies employed in this study promotes more out-of-layer side-branching of the solidification cells, resulting in a significant broadening of the grains, hence reducing their columnarity [40]. This explains why the meander 0° scan strategy induced the most columnar microstructure and the smallest grain size (Figure 3 and Figure 4), in particular on the section perpendicular to the build direction. Micro-plasticity in front of a crack tip affects the crack opening, and thereby the propagation rate of short fatigue cracks. Grain boundaries are known to restrict dislocation slip ahead of the crack tip, and hence hinder crack propagation. Therefore, the high number of grain boundaries that were perpendicular to the crack propagation direction can significantly slow crack advancement in the meander 0° sample; resulting in the influential role of the scan strategy on the fatigue crack propagation rate.

## 5. Conclusion

This study discusses significant new insights into the relationship between process conditions (in particular scan strategy and surface finish), consolidation, solidification microstructure and fatigue behaviour of the CrMnFeCoNi high-entropy alloy fabricated by laser powder-bed fusion. Three considered printing strategies resulted in columnar grains, which consisted of fine solidification cells. The degree of columnarity was found to be governed by the scan strategy. STEM-EDX analyses showed that Cr, Co and Fe segregated to the cell core during solidification, while Mn and Ni segregated to cell boundaries, in agreement with CALPHAD predictions. The dense dislocation tangles located at the cell boundaries restricted dislocation slip during plastic deformation, resulting in an improved yield stress compared to cast CrMnFeCoNi HEA. Dislocation slip and deformation twinning were the main deformation mechanisms of the HEA in monotonic loading. However, dislocation slip was the single dominant deformation mechanism responsible for cyclic plasticity. Key-hole porosity in sub-surface regions was seen to act as preferential sites for fatigue crack initiation in the as-printed samples, while lack-of-fusion defects were major initiation sites for surface-finish condition. Machining significantly improved the fatigue life of the alloy by removing sub-surface key-hole pores and hence delaying fatigue crack initiation. The print strategies were not found to have a significant influence on the fatigue lives of the HEA samples. However, print strategies were found to have considerable effects on fatigue crack propagation. In particular, the meander 0° sample exhibited higher resistance to fatigue crack propagation compared to the other considered strategies, since this strategy induced the smallest spacing of grain boundaries along the crack propagation direction.

# Acknowledgements

We acknowledge the financial support provided by Engineering and Physical Sciences Research Council (ESPRC) [Grant EP/K503733/1] and the Excellent Funds for Frontier Research at Imperial College London